\documentclass[12pt,a4paper]{article}
\usepackage{natbib} 
\usepackage{multirow,bm}
\usepackage[latin1]{inputenc}
\usepackage{authblk}    
\usepackage{makeidx}         
\usepackage{graphicx}        
\usepackage{amsmath}
\usepackage{amsthm}
\usepackage{amsfonts}
\usepackage{appendix}
\usepackage{url}
\usepackage{subcaption} 
\usepackage{algorithm}
\usepackage{algpseudocode}
\usepackage{adjustbox} 
\usepackage{bbm}
\usepackage[colorlinks=true, allcolors=blue]{hyperref}
\newcommand{\mailadd}{\href{mailto:me@somewhere.com}{\vspace*{0.15cm}}} 



\title{\textbf{Joint modelling of national cultures accounting for within and between-country heterogeneity}}

\author[a]{Veronica Vinciotti}
\author[b]{Luca De Benedictis}
\author[c]{Ernst C. Wit}
\affil[a]{\small University of Trento; \mailadd{\texttt{veronica.vinciotti@unitn.it}}}
\affil[b]{\small University of Macerata}
\affil[c]{\small Universit\`a della Svizzera italiana}
\date{}
\begin{document}
\maketitle
\begin{abstract}
Cultural values vary significantly around the world. Despite a large heterogeneity, similarities across national cultures are present. This paper studies cross-country culture heterogeneity via the joint inference of country-specific copula graphical models from world-wide survey data. To this end, a random graph generative model of the cultural networks is introduced, with a latent space and proximity measures that embed cultural relatedness across countries. Within-country heterogeneity is also accounted for, via parametric modelling of the marginal distributions of each cultural trait. All together, the different components of the model are able to identify several dimensions of culture.
\end{abstract}


\section{Introduction}
\label{intro}
Culture, broadly defined by a set of beliefs, values, norms, and practices \citep{cuche2020notion}, plays a fundamental role in shaping societal frameworks across the globe \citep{alexander1990culture}. It not only influences individual behaviour but also has broader implications for economic policies, political dynamics, and international relations. Given its pivotal role, there is an increasing academic interest in cultural studies as a means to understand the nuances of global interactions in an era of sustained countries' interdependence. While the qualitative exploration of culture has provided depth and context, the increasing availability of cross-country data and the opportunity to adopt a comparative perspective of national cultures underscore the need for a quantitative approach. Indeed, quantification of cultural aspects, using well-defined statistical methodologies, 
allows for an objective, replicable and scalable study of individual national cultures and comparative cultural analyses. 

International surveys, such the ones used in international business and organisational psychology \citep{hofstede1984cultures, hofstede2011dimensionalizing}, international management \citep{house2004culture} or the European Values Study and the World Values Surveys (EVS/WVS)  used in the present study \citep{Ing2018}, are a rich source of data for quantitative studies of national cultures. For example, \cite{acemoglu2021successful} use EVS/WVS data in a recent cross-country study on democracy. The questions that are contained in these surveys reflect the views of the population on various aspects, from abortion, homosexuality to the importance of religion, and can thus be taken to represent a latent dimension of culture. 

In a recent study, \cite{debenedictis23} argue how national cultures cannot be described solely by the response of people to individual survey questions. Statistically, marginal distributions of \textit{cultural traits} depict only part of a national culture. The interconnectedness of these traits can in itself be an interesting feature of culture. Thus, statistical quantification of national cultures should be based both on the marginal distributions of the cultural traits and on the \textit{cultural networks} that describe the statistical dependence between these cultural traits. \cite{debenedictis23} show that the Gaussian copula graphical models provide an appropriate statistical framework to couple both the information from the marginals and their statistical dependence into one model of national culture. The aim of the current paper is to look more closely at cross-country cultural comparisons and three statistical challenges that arise from that. 

The first aspect is that there is typically heterogeneity in the population of respondents to a survey. If a demographic characteristic of the respondent, say gender, has an effect on their view in matter of culture, then the marginal distribution of specific cultural traits may be different when split between, say, male and female respondents. In this case, a statistical adjustment is needed to avoid introducing sources of bias in the follow-up analysis. In particular, ignoring this aspect could have an impact both in the statistical inference of the cultural networks, as these are typically less affected by the demographic characteristics of the respondent than the marginal distributions, and in the cross-country comparisons, as different countries will most likely have different sample configurations.  
This paper addresses this challenge with the use of parametric regression models within a Gaussian copula graphical modelling framework, as advocated also in other applications \citep{vinciotti22}. In particular, ordinal regression models will be used to link the response to a specific survey question to characteristics of the respondent, such as age and gender.

The second aspect is that some structural similarity is to be expected between the cultural networks associated to the different countries. For example, \cite{debenedictis23} show how tolerance towards abortion and homosexuality are strongly and positively associated across all countries considered in the previous study. Other similarities may be more local, e.g., only between groups of countries. While this was observed and quantified a posteriori, i.e., after the inference of the individual national cultures, this paper argues that this information should in fact be embedded in the model. Indeed, a) similarities between countries aid statistical inferences of the individual networks when using a joint statistical model, and b) similarity and heterogeneity between national cultures can be better quantified by a joint statistical model. This paper proposes a random graph model for the joint distribution of the conditional independence graphs associated to the Gaussian copula graphical models of each country. The resulting \textit{random graphical model} \citep{vinciotti23} is defined via a latent space, that allows to identify relatedness across the different countries at the structural level. Proximity between countries in this space, which is inferred from data, is associated to structural similarity between the cultural networks. 

Beyond assessing the extent of cross-country cultural heterogeneity, an important aim of this paper, and the third statistical aspect that is considered, is that of explaining cultural heterogeneity. To this end, the proposed random graphical model is augmented with the inclusion of potential drivers of cultural heterogeneity, such as geographical distance and past historical connections between the countries as revealed by language commonality.  These are indeed expected to play a key role in explaining the location of countries in the cultural spectrum, in line with the role that they have been found to play in social, political and economic studies \citep{de2011gravity, head2014gravity, yotov2022gravity}.

Combining the three aspects into one statistical model requires integration of data at various levels and from various sources. On one hand, there is the EVS/WVS data on 10 selected questions (mapped to 10 cultural traits) from 84 countries around the world, with, on average, 1740 respondents in each country. This is the primary source of data for the country-specific Gaussian copula graphical models. These data are jointly modelled via the latent space of the random graph model, which generates structural dependences between cultural traits across countries. In addition to these, the model is able to integrate data from external sources both at the node level, namely the respondents' age and gender in the marginal distributions of each cultural trait, and at the edge levels, namely geographical proximity and linguistic commonality in the random graph generative model. The Bayesian inferential procedure that is developed is able to fully account for the statistical uncertainty at all the individual components of the joint model. 

The paper is structured as follows. Section \ref{data} describes the different sources of data that will be considered. Section \ref{rgm} describes the proposed random graphical model, while Section \ref{culture} shows how the proposed model allows to measure the extent of cross-country cultural heterogeneity and to identify its potential drivers. Finally, Section \ref{conclusion} draws conclusions and suggests directions for future work.

\section{Integrating different sources of data} \label{data}
\subsection{Survey data on cultural values} 
 \begin{figure*}[!t]
\centering
 \includegraphics[width=\textwidth]{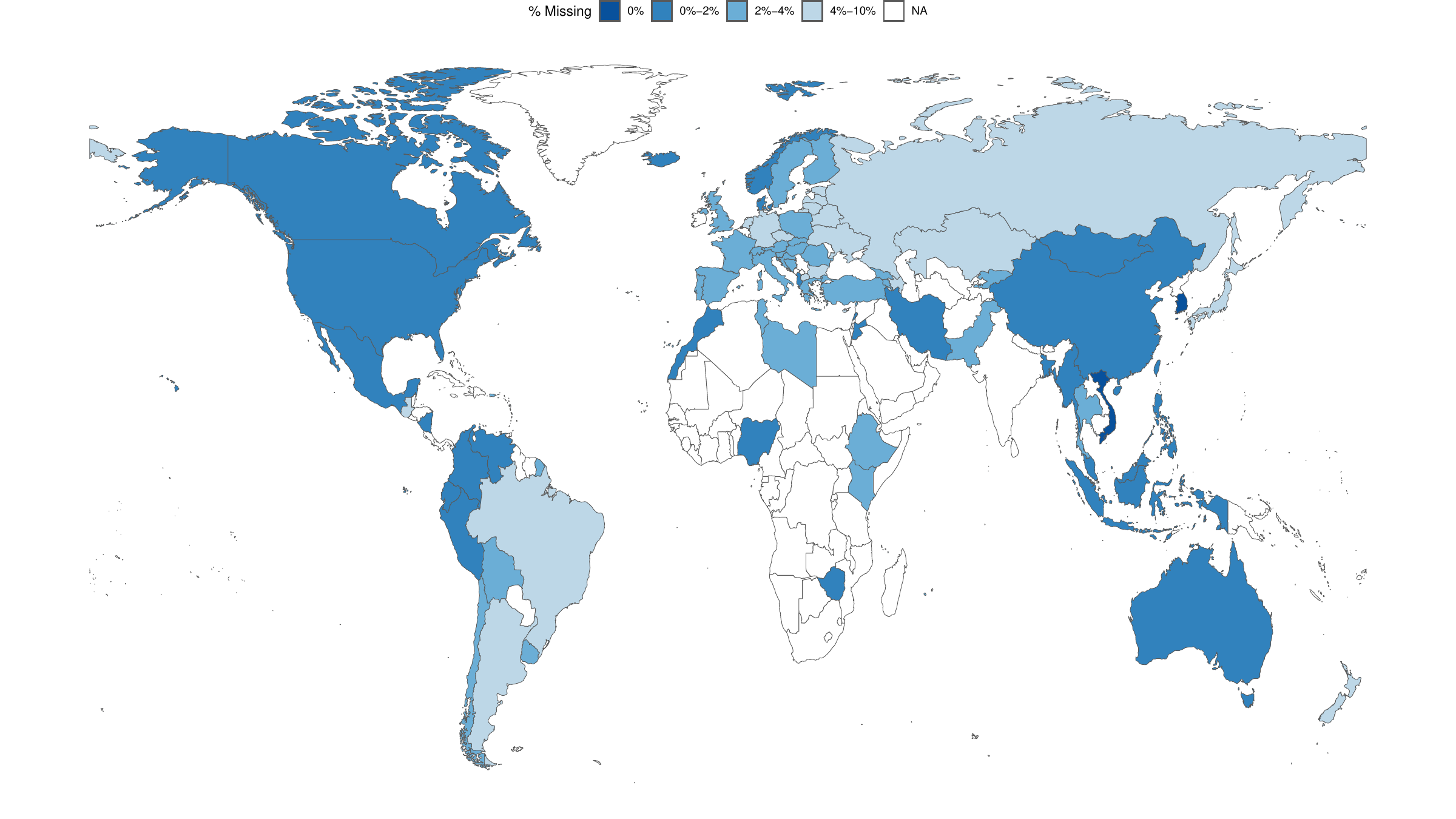}
	\caption{Coverage (in blue) of the 84 countries considered in this study from the wave 7 of the EVS/WVS survey. Different shades of blue correspond to 4 different levels of missingness, measured as an average over the 10 cultural traits. The countries not in the study are in white.}
  \label{fig:ewvs7map}
	\end{figure*}
The primary source of data is given by responses to survey questions. We focus in particular on the joint EVS/WVS survey, which covers 90 countries around the world and contains a large number of questions that are common to the EVS and WVS surveys \citep{evswvs22}. As in \cite{debenedictis23}, and following from the tradition of \cite{IngWel2005}, we select the same 10 cultural traits considered in these studies and collect responses to these questions from the latest wave of the survey (Wave 7, 2017-2021). Figure \ref{fig:ewvs7map} shows in blue the 84 countries for which there is data for all 10 selected cultural traits.

Table \ref{tab:culturaltraits} provides a description of the 10 cultural traits, and the corresponding question or index associated to each of them in the EVS/WVS survey.  
\begin{table}[!t]
\caption{Descriptive statistics of survey data: Column id includes the joint EVS-WVS Wave 7 questionnaire codes; column  \texttt{variable} contains a longer description of the variable and specifies the number of categories and their direction if ordinal;  column 3 and 4 contain the overall average value of each variable and the percentage of missing values, while the [min,max] intervals refer to the lowest/highest values of mean and \% of missing, respectively, across the 84 countries.  \label{tab:culturaltraits}}
\centering
\begin{tabular}{llcc} \hline
id     & \texttt{variable} (categories, direction, label)          &  overall mean      & missing \%\\
      &             &  [min, max]       & [min, max] \\[0.5ex] \hline 
\multicolumn{4}{l}{\textbf{Survey data on cultural traits}} \\  \hline
\multirow{2}*{Q46}  &\multirow{2}*{level of \texttt{happiness} (1:4, \texttt{low}$\rightarrow$ \texttt{high}, \texttt{H})}        & 1.9           & 0.9          \\[-1ex]
                       &   & [1.4, 2.5]           & [0.0, 11.2]    \\[0.5ex]
\multirow{2}*{Q57} & \multirow{2}*{\texttt{trust} in people  (1:2, \texttt{high}$\rightarrow$ \texttt{low}, \texttt{T})}          & 1.7       & 1.7            \\[-1ex]
                &                     & [1.2, 2.0]   & [0.0, 5.9]    \\[0.5ex]

\multirow{2}*{Q49} & \multirow{2}*{\texttt{respect for authority} (1:3, \texttt{high}$\rightarrow$ \texttt{low}, \texttt{R}) }          &  1.6          & 4.2                 \\[-1ex]
                &                           & [1.1, 2.8]          & [0.00, 18.8]                  \\[0.5ex]

\multirow{2}*{Q209}  & \multirow{2}*{\texttt{voice} through petitions (1:3, \texttt{high}$\rightarrow$ \texttt{low}, \texttt{V}) }        & 2.1            & 3.3         \\[-1ex]
                &          &  [1.2, 2.7]          & [0.0, 13.9]    \\[0.5ex]

\multirow{2}*{Q164}  & \multirow{2}*{\texttt{importance of God} (1:10, \texttt{low}$\rightarrow$ \texttt{high}, \texttt{G}) }           & 6.8       & 1.7            \\[-1ex]
                &                  & [2.8, 9.9]   & [0.0, 7.9]    \\[0.5ex]

\multirow{2}*{Q182} & \multirow{2}*{justification of \texttt{homosexuality} (1:10, \texttt{low}$\rightarrow$ \texttt{high}, \texttt{O}) }         & 4.5            &  3.8                \\[-1ex]
                &                           & [1.1, 9.0]           & [0.00, 14.5]                \\[0.5ex]

\multirow{2}*{Q184}  & \multirow{2}*{justification of \texttt{abortion} (1:10, \texttt{low}$\rightarrow$ \texttt{high}, \texttt{A})}          & 4.3            & 2.8          \\[-1ex]
                &            & [1.4, 8.4]            & [0.0, 11.2]    \\[0.5ex]

\multirow{2}*{Q254}  & \multirow{2}*{\texttt{national pride} (1:4, \texttt{high}$\rightarrow$ \texttt{low}, \texttt{P})}           & 1.6       & 4.3           \\[-1ex]
                &                     & [1.1, 2.2]   & [0.0, 52.5]    \\[0.5ex]

\multirow{2}*{Y002} & \multirow{2}*{\texttt{post-materialism} (1:3, \texttt{low}$\rightarrow$ \texttt{high}, \texttt{M})}         & 1.9           & 3.9                 \\[-1ex]
                &                            & [1.5, 2.3]         & [0.0, 33.5]                \\[0.5ex]

\multirow{2}*{Y003}  & \multirow{2}*{\texttt{obedience vs independence}   (1:5, \texttt{high}$\rightarrow$ \texttt{low}, \texttt{B})}        &  3.3          &  2.1\\[-1ex]
                &                         & [2.2, 4.2]         & [0.0, 50.0]  \\\hline
\multicolumn{4}{l}{\textbf{Characteristics of survey respondents}} \\ \hline
\multirow{2}*{Q260} & \multirow{2}*{\texttt{gender} (1: M, 2: F)} & 1.5 & 0.08 \\[-1ex]
                &                         & [1.5, 1.7 ]         & [0, 2.2]  \\[0.5ex]
\multirow{2}*{Q262}  & \multirow{2}*{\texttt{age} (continuous)} & 45.9 & 0.6\\[-1ex]
                &                         & [30.7, 57.7]         & [0.0, 20.5]  \\\hline
\end{tabular}
\end{table}
All variables are ordinal, and capture various aspects and views of the respondent, namely their level of happiness, trust in others, respect for a greater authority, availability to express personal opinions by signing petitions,  importance of religion, justification of homosexuality or abortion, and pride to be a citizen of a certain country. The last two traits are indices and are obtained from a combination of questions: post-materialism is a composite index associated to how much one sees happiness on material things and economic stability versus freedom, creativity, self-expression, autonomy; while, obedience/independence measures the importance to teach children to have religious faith and to obey, rather than to be independent and to pursue perseverance and determination. 

The summary statistics from Table \ref{tab:culturaltraits} are generally in line with these from the \cite{debenedictis23} study, which is however covering the previous wave and 54 countries, of which only 43 are in common with the current study. The level of missingness is generally low, less than 10\% on average across all cultural traits, as visualized for each country in Figure \ref{fig:ewvs7map}. However, there are some noticeable exceptions, namely the variable \texttt{obedience vs independence} with 50\% of missing data in Cyprus and the variable \texttt{national pride} with 52.5\% of missing in Andorra. The graphical modelling approach described in Section \ref{rgm} is able to handle missing data without any need for prior removal or imputation. Under an assumption of missing-at-random, each missing value is projected via a Gaussian copula to the full interval of real values during MCMC sampling of the latent Gaussian data.

\subsection{Data on survey respondents' characteristics}
A second source of data refers to the demographic characteristics of the respondents, which may have an effect on their views towards culture. Among those that are collected as part of the EVS/WVS data and sufficiently observed across all countries, we look closely at the age and gender of the respondents.

\begin{figure}[!tb]
	\centering
	\includegraphics[width=0.49\textwidth]{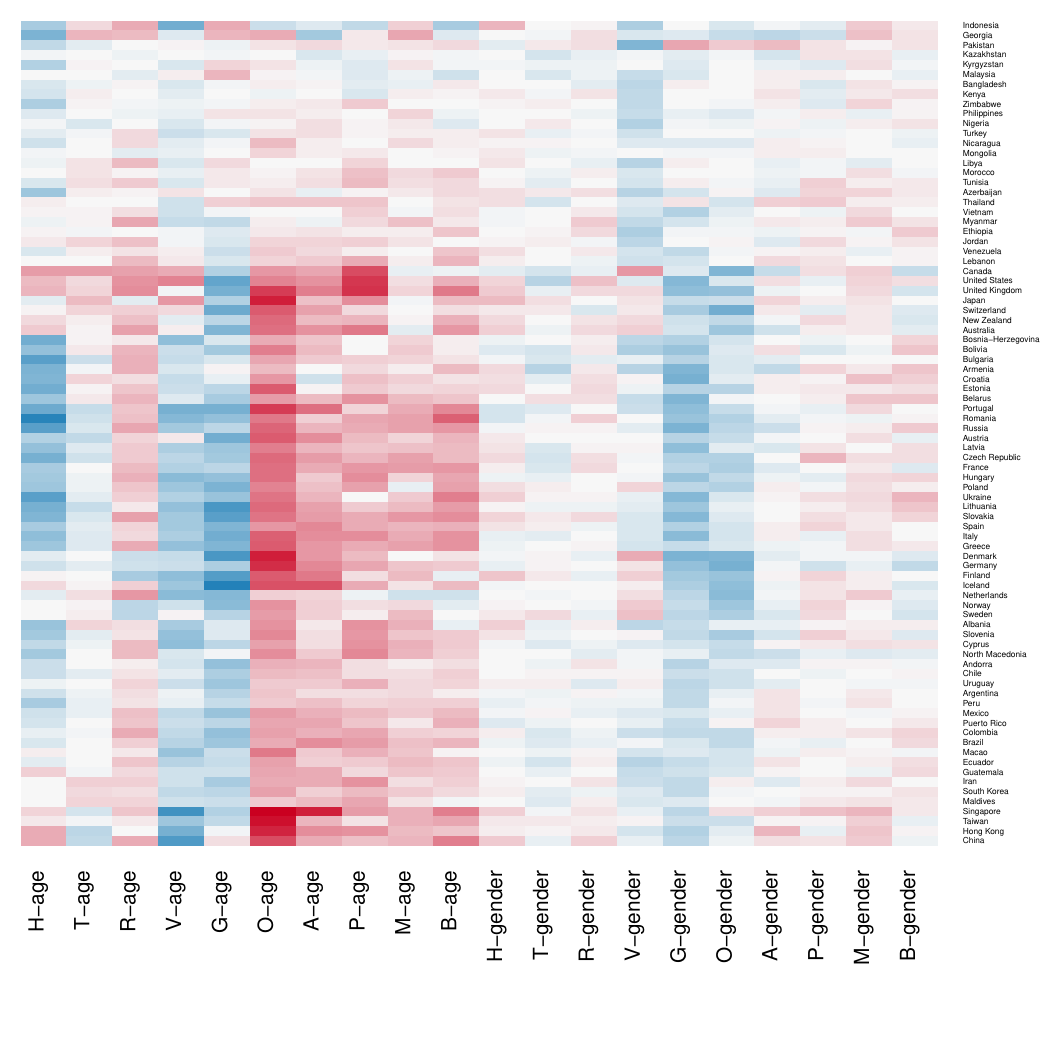}
	\caption{Standardized regression coefficients of age and gender, from an ordinal regression model for each cultural trait and country. The colour scale goes from negative (red) to positive (blue).}
	\label{fig:betacoef}
\end{figure}

Figure \ref{fig:betacoef} reports the standardised coefficients for age and gender, obtained from an ordinal regression model for each cultural trait and country, which will be more formally defined in Section \ref{rgm}. A positive (blue) coefficient represents a positive association with the ordinal variable, while negative (red) refers to a negative association.

With some variability across countries, the plot shows how age tends to be more strongly associated with the cultural traits than gender. In particular, looking at the most significant effects and considering the ordering of the variables in Table \ref{tab:culturaltraits}, the data suggest that the older people are, the more they tend to be happy, the less they tend to express their opinion by signing petitions, the more religious they are, the less tolerant they are towards homosexuality and abortion, the more proud they are of their nationality, the more materialistic they are, and the more they teach obedience to their children. As for gender, the more pronounced effect across countries appears to be in terms of importance of religion and justification of homosexuality. In particular, women, while tending to be more religious, also tend to be more tolerant towards homosexuality than men.

\begin{figure*}[!tb]
	\centering
	\begin{tabular}{cc}
	\includegraphics[width=0.4\textwidth]{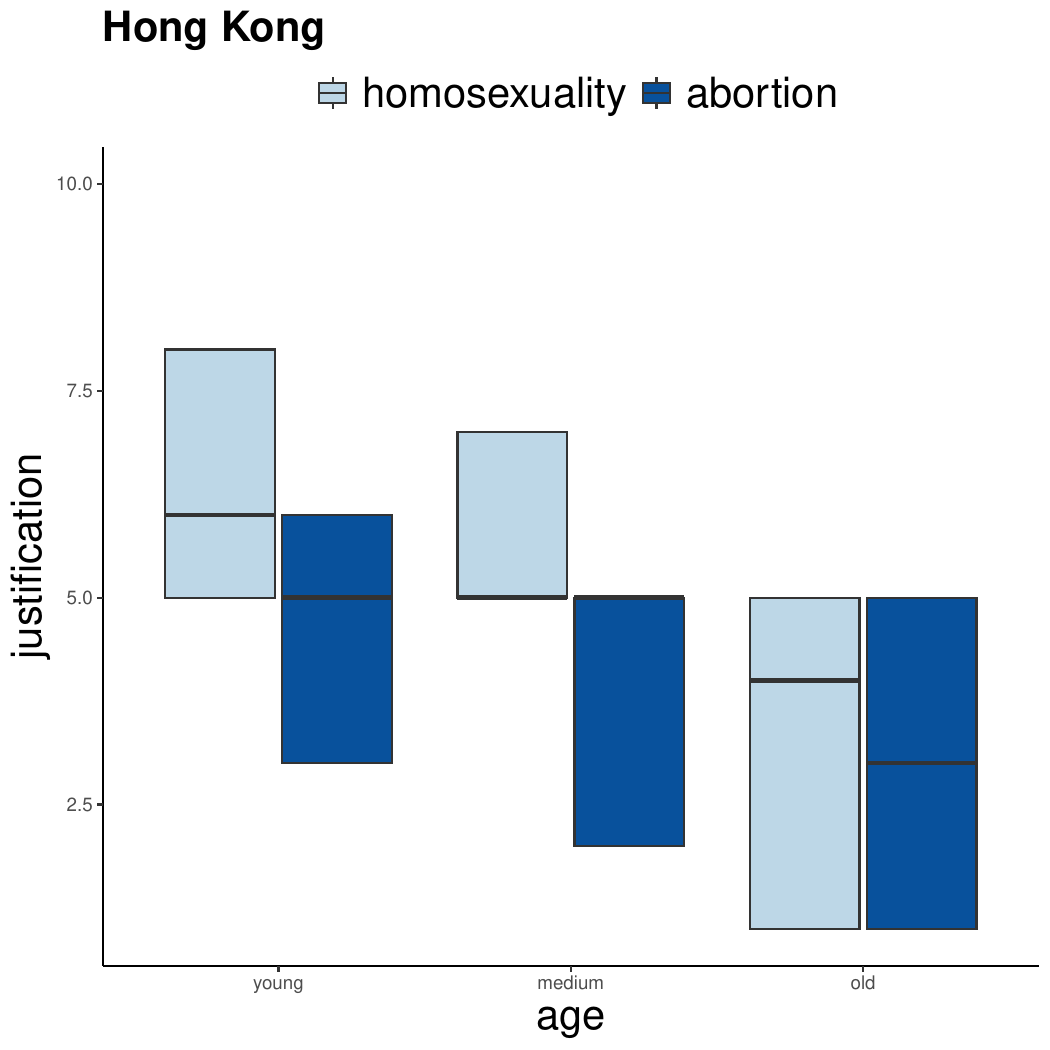}&
	\includegraphics[width=0.4\textwidth]{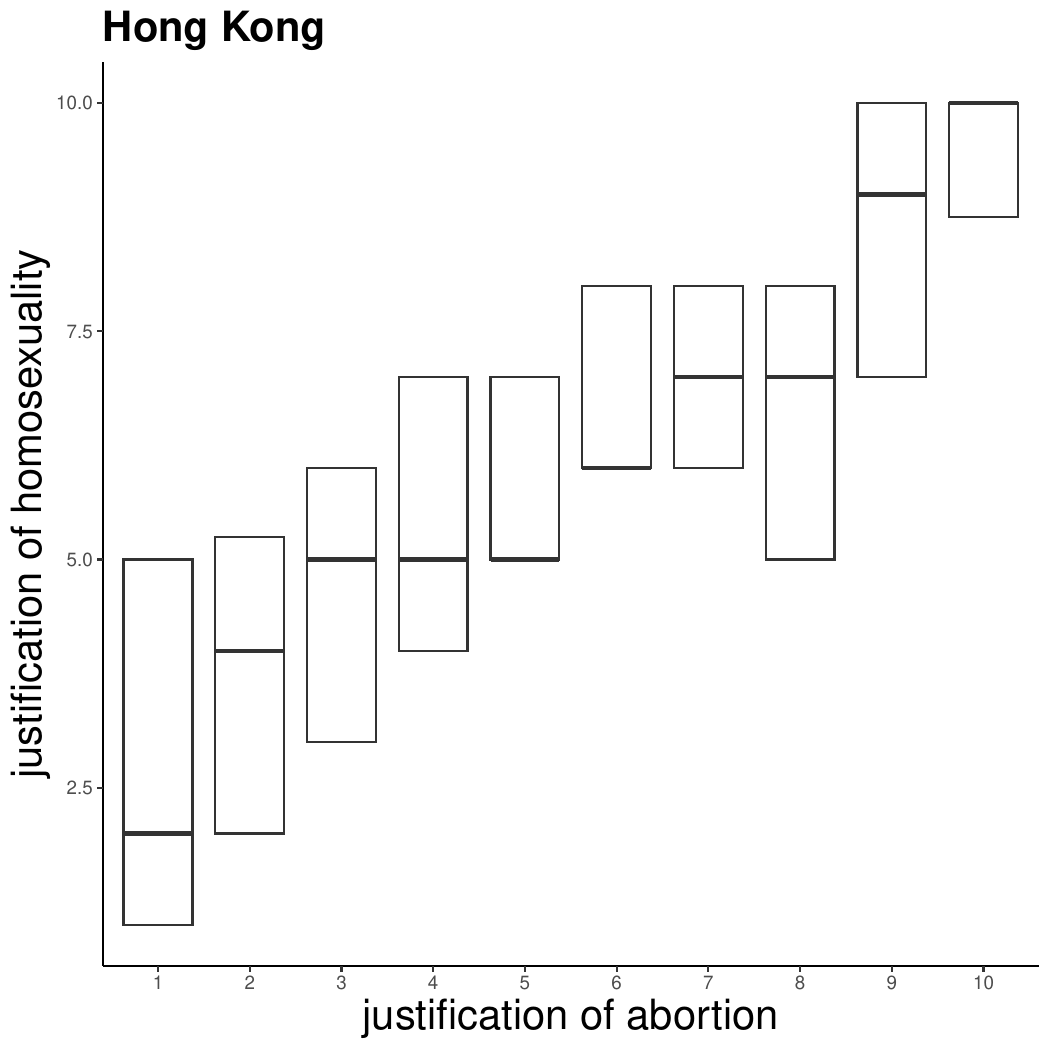} 	 \\
	(a) & (b) \\
	\includegraphics[width=0.49\textwidth]{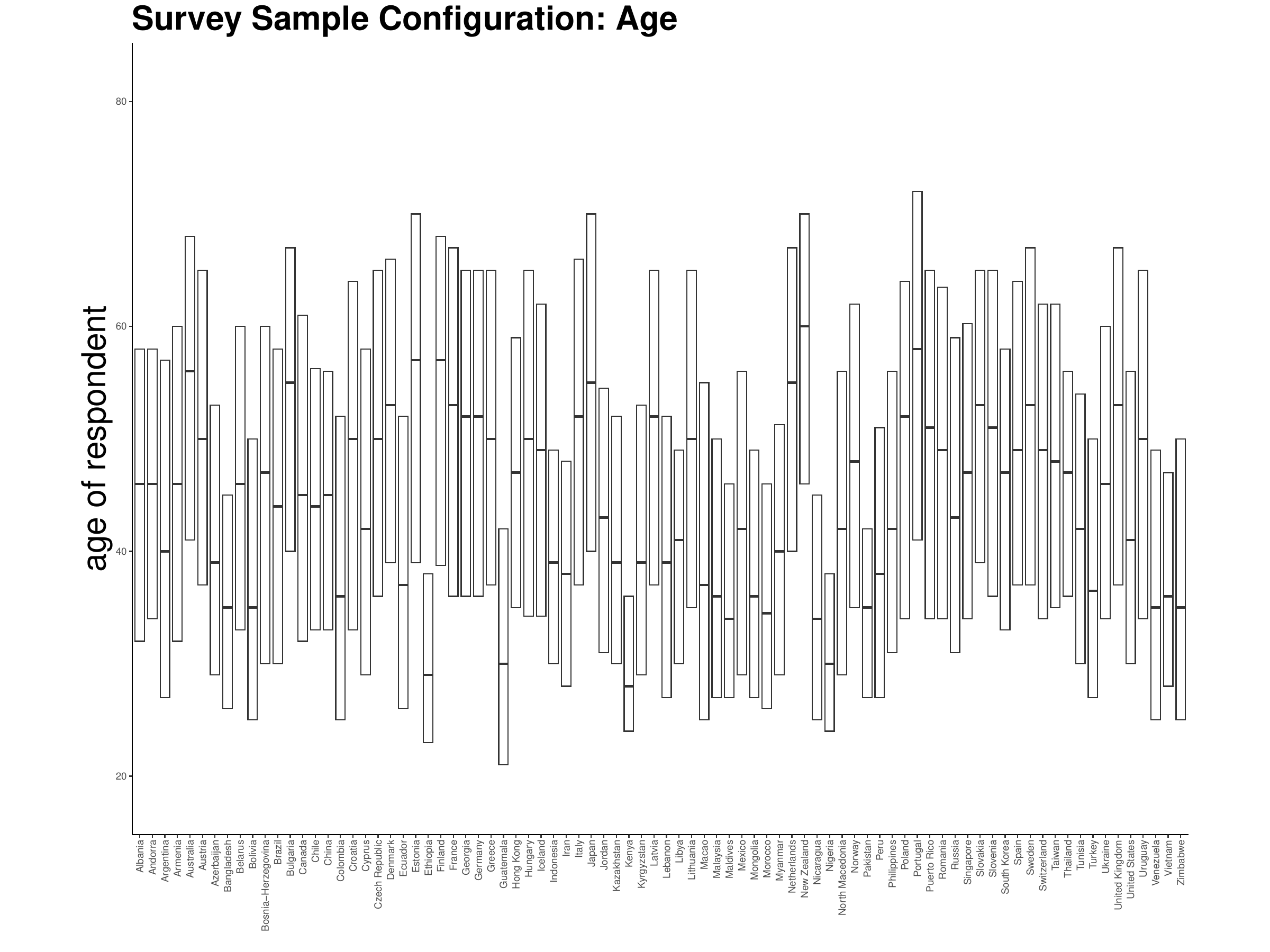} & 
	\includegraphics[width=0.49\textwidth]{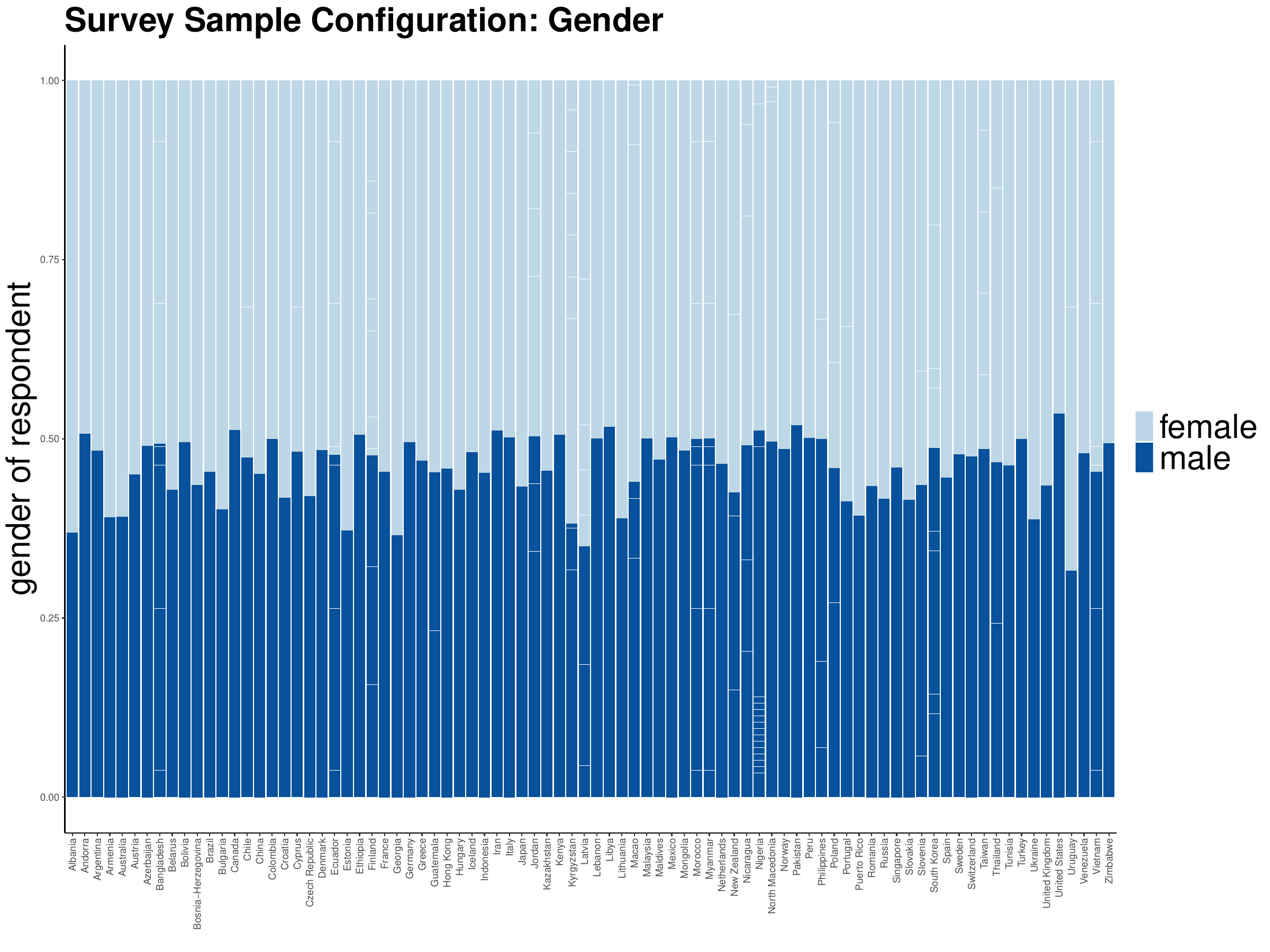}\\
	(c) & (d)
	\end{tabular}
	\caption{a) According to the survey results in Hong Kong, tolerance towards homosexuality and abortion are positively correlated, b) and both decrease with age; (c) distribution of age and d) gender throughout the entire survey.}
	\label{fig:marginalsHK}
\end{figure*}

The top panel of Figure \ref{fig:marginalsHK} focusses on one of the countries in the sample, Hong Kong. Figure \ref{fig:marginalsHK}a confirms how age has a strong effect on the tolerance of people towards homosexuality and abortion at the marginal level, with older people being less tolerant than the younger ones. Figure \ref{fig:marginalsHK}b shows how the two cultural traits have a strong positive association. While the level of correlation may not depend strongly on age (as an indication, correlation is 0.38 for young people, 0.49 for middle-aged people, and 0.42 for elderly people), the marginal effects may distort the estimation of the overall correlation between these two cultural traits. Even if there is no correlation between two cultural traits, then the marginal differences in, say, age groups will in fact lead to an overall apparent correlation. This will not only bias the measurement of dependence between cultural traits in Hong Kong, but it will also distort comparative studies between countries with different marginal effects or different sample configurations. The lower panel of Figure \ref{fig:marginalsHK} shows how the latter is indeed the case, particularly for age (Figure \ref{fig:marginalsHK}c), with New Zealand the country with the oldest survey respondents (mean equal to 57.7), while Kenya the one with the youngest survey respondents (mean equal to 30.7) and indeed showing no marginal effects for age in Figure \ref{fig:betacoef}.


From this exploratory analysis, we will develop a model that includes a marginal adjustment for age and gender for each country and cultural trait, while the dependence structure between cultural traits will be assumed to be only country-specific.

\subsection{Data on country similarities} \label{sec:distances}
A third source of data is characterized by measures of proximity between countries that may be a potential driver of cultural heterogeneity. These are extracted from the CEPII gravity database, which is often used for economic studies (see \cite{conte22} and references therein). In particular, we will consider the following variables:
\begin{itemize}
\item Geographical proximity: 1/log(distance), where the distance is taken as the distance between the most populated city of each country in km \citep{disdier2008puzzling};
\item Sharing of primary spoken language: 1 if the two countries share a common official or primary language, 0 otherwise \citep{melitz2014native}; 
\item Sharing of major spoken language: 1 if the two countries share a common language spoken by at least 9\% of the population, 0 otherwise \citep{melitz2014native};
\item Same continent: 1 if the two countries belong to the same continent, 0 otherwise \citep{disdier2008puzzling}.
\end{itemize}
In the next section, we describe a joint model of national cultures that exploits the information from the three sources of data described above to account both for within and between-country heterogeneity.

\section{Random graphical model of cultural heterogeneity}
\label{rgm}

In this section, we define a \textit{random graphical model} \citep{vinciotti23} for describing cultural networks that induce survey data for each country, while accounting for 1) similarities between countries at the level of cultural networks, 2) potential driving factors of cultural heterogeneity and 3) demographic characteristics of the respondents.  For country  $k=1,\ldots, K$, let $\mathbf{Y}^{(k)}=(Y^{(k)}_1,\ldots,Y^{(k)}_p)$ be the random $p$-dimensional vector of the responses to $p$ survey questions. In our study, there are $K=84$ countries and $p=10$ survey questions.  

Since the responses to the questions are ordinal, we consider a Gaussian copula graphical model for each $\mathbf{Y}^{(k)}$. Namely,
\begin{align*}
&P(Y^{(k)}_{1} \leq y_{1},\ldots, Y^{(k)}_{p} \leq y_{p}) = \Phi_{\boldsymbol{\Omega}^{(k)}} \big( \Phi^{-1}(F^{(k)}_{1}(y_1)), \ldots, \Phi^{-1}(F^{(k)}_{p}(y_p)) \big),
\end{align*}
where $\Phi_{\boldsymbol{\Omega}^{(k)}}$ is the cumulative distribution function of a $p$-dimensional multivariate normal with a zero mean vector and precision matrix $\boldsymbol{\Omega}^{(k)}$, $\Phi$ is the standard univariate normal distribution function, and $F^{(k)}_{j}$ is the marginal distribution of cultural trait $j$ in country $k$. The dependence structure induced by this model in condition $k$ is represented by the conditional independence graph $G^{(k)}$. Following from the theory of Gaussian graphical models \citep{lauritzen96}, this is given by the zero-patterns of the precision matrix $\boldsymbol{\Omega}^{(k)}$. 

Given the heterogeneity of the population of respondents, we account for covariates, such as age and gender, by modelling the marginals $F^{(k)}_{j}$ parametrically. In particular, we consider ordinal regression models for each country and each cultural trait. Namely,
\begin{align}
\label{eq:marginals}
F^{(k)}_{j}(c|\bm{X}=\bm{x}) =\eta^{(k)}_{jc}+\boldsymbol{\gamma}^{(k)}_j\bm{x}
\end{align}
with thresholds $\eta^{(k)}_{jc}$ for category $c$ of cultural trait $j$ in country $k$, and regression coefficients $\boldsymbol{\gamma}^{(k)}_j$ associated to node covariates $\bm{x}=(1,x_1,\ldots,x_m)^\top$. 

As for the random graph model, describing the joint distribution of the cultural networks $G^{(k)}$, we are particularly interested in modelling the relatedness of the different countries as well as a possible association with potential drivers. To this end, we consider the following latent probit network model \citep{hoff02}
\begin{align}
\label{eq:latentprobit}
P({G_{j_1,j_2}}^{(k)}=1~|~G_{j_1,j_2}^{(-k)}, \mathbf{sim})&=  \Phi\Big(\alpha_k +\boldsymbol\beta^t\sum_{k' \ne k}\mathbf{sim}_{kk'}(1_{\{{G_{j_1,j_2}}^{(k')}=1\}}-1_{\{{G_{j_1,j_2}}^{(k')}=0\}}) \notag\\
&+\mathbf{c}_k^t\sum_{k' \ne k}\mathbf{c}_{k'}(1_{\{{G_{j_1,j_2}}^{(k')}=1\}}-1_{\{{G_{j_1,j_2}}^{(k')}=0\}})\Big),
\end{align}
where 
${G_{j_1,j_2}}^{(k)}=1$ defines an edge between node $Y_{j_1}$ and node $Y_{j_2}$ in country $k$, $\mathbf{sim}_{kk'} \in \mathbb{R}^d$ is a vector of proximity measures between country $k$ and country $k'$, such as geographical proximity,  $\boldsymbol\beta$ is the corresponding $d$-dimensional vector of parameters,   $\mathbf{c}_k\in \mathbb{R}^2$ is the latent space vector of parameters for country $k$ and determines its location in the latent space, $\alpha_k$ is the intercept of the model and relates to the overall sparsity level of graph  $G^{(k)}$. Given the model formulation, an edge between cultural trait $Y_{j_1}$ and cultural trait $Y_{j_2}$ in the cultural network $G^{(k)}$ of country $k$ is more (less) likely if that country is close to a country $k'$ where that edge is present (missing). Vicinity can be both in terms of proximity measures, in which case the probability is tuned via the parameters $\boldsymbol\beta$, and in terms of the locations of the countries in the latent space, in which case it is measured by the inner product of the  $\mathbf{c}_k$ and $\mathbf{c}_{k'}$ parameters.

Bayesian inference is conducted for the full set of model parameters, namely $\eta^{(k)}_{jc}$, $\boldsymbol{\gamma}^{(k)}_{j}$ at the marginal level, while $G^{(k)}$, $\boldsymbol{\Omega}^{(k)}$, $\alpha_k$, $\mathbf{c}_k$ and $\boldsymbol{\beta}$ at the structural level, with $k=1,\ldots, K$ countries and $j=1,\ldots,p$ cultural traits. The three components of the joint model, namely country-specific regression models for each marginal trait, country-specific Gaussian copula networks, latent probit model of the cultural networks from all countries, are joined together in the inferential procedure. In particular, similar to \cite{vinciotti23}, parameter estimation is based on the following steps:
\begin{enumerate}
\item Fitting of the $\eta^{(k)}_{jc}$, $\boldsymbol{\gamma}^{(k)}_{j}$ marginal parameters in Equation (\ref{eq:marginals}), leading to the intervals
\begin{equation*}
\mathcal{I}(y^{(k)}_{ij}|\bm{x^{(k)}_{i}})  = \big(\Phi^{-1} \big(F^{(k)}_j(y^{(k)}_{ij}-1|\bm{x}^{(k)}_{i}) \big),\Phi^{-1} \big(F^{(k)}_j(y^{(k)}_{ij}|\bm{x}^{(k)}_{i}) \big)\big],
\end{equation*}
with $y^{(k)}_{ij}$ the response to question $j$ by individual $i$ in country $k$, and $\bm{x}^{(k)}_{i}$ its vector of covariates (age and gender);
\item Based on the current graphs $G^{(k)}$ for all countries, Gibbs sampling of the parameters $\alpha_k$, $\mathbf{c}_k$ and $\boldsymbol{\beta}$, $k=1,\ldots,K$, of the latent network probit model in Equation (\ref{eq:latentprobit});
\item Truncated on the intervals from step 1, Gibbs sampling of Gaussian random vectors for each observation in each country, based on the current estimates of the precision matrices $\boldsymbol{\Omega}^{(k)}$;
\item Conditional on the latent Gaussian data from step 3, Gibbs sampling of precision matrices $\boldsymbol{\Omega}^{(k)}$ from their posterior G-Wishart distributions;
\item Continuous time birth-death MCMC sampling of the next graphs, by adding (birth) or deleting (death) an edge from the current graphs $G^{(k)}$ for each country, with the birth and death rates informed by the current estimates of the latent network probit model from step 2. The last three steps conclude the sampling from the Gaussian copula, before returning to step 2, and iterating these steps until convergence.
\end{enumerate}

\section{Explaining cultural heterogeneity} \label{culture}
We now use the data described in Section \ref{data} to study the inter-relatedness of various socio-cultural dimensions around the world. In previous work, \cite{debenedictis23} fit Gaussian copula graphical models for each country separately and observe how the distance between countries in terms of their cultural values depends both on differences between the marginal responses to the cultural traits and between the dependence structure of these cultural traits, i.e., the cultural networks.  By embedding structural similarities between cultural networks within a joint modelling framework, the random graphical model proposed in this paper allows us to consider the second aspect more in depth. In particular, firstly, we will use it to measure the extent of cultural heterogeneity between the cultural networks and the relative similarities between countries in the cultural spectrum. Secondly, we identify potential drivers of this cultural heterogeneity.

\subsection{Exploring cultural heterogeneity}
\label{latentspace}
We first explore the extent of cultural heterogeneity around the world by fitting a latent space model without any of the drivers. In particular, we model the marginals as in (\ref{eq:marginals}), and consider a random graph generative model defined by
\begin{align}
\label{eq:modelLS}
& P({G_{j_1,j_2}}^{(k)}=1~|~G_{j_1,j_2}^{(-k)})=  \Phi\Big(\alpha_k +\mathbf{c}_k^t\sum_{k' \ne k}\mathbf{c}_{k'}(1_{\{{G_{j_1,j_2}}^{(k')}=1\}}-1_{\{{G_{j_1,j_2}}^{(k')}=0\}})\Big).
\end{align}
That is, the marginals capture the country-specific effects on culture from the individual responses to the survey questions, adjusted by age and gender, while the latent space accounts for the full heterogeneity in the cultural network component. So the vicinity of countries in this latent space is associated to structural similarities in the corresponding cultural networks, and viceversa for distant countries.

We fit the model on the latest EVS/WVS data for $K=84$ countries and $p=10$ cultural traits, as described in Section \ref{data}, adjusting for age and gender at the marginal level. For the latter, we use the \texttt{polr} function in R to fit the ordinal regression models in (\ref{eq:marginals}) and extract the regression coefficients $\boldsymbol{\eta}_j$ and $\boldsymbol{\gamma}_j$ for $j=1,\ldots,p$. Figure \ref{fig:betacoef}, already discussed in Section \ref{data}, is obtained from the $\hat{\boldsymbol{\gamma}_j}$ of these models. For the Bayesian structural learning sampling described in the previous section, we set only weakly informative $N(0,10)$ priors on each parameter of the random graph model, namely $\alpha_k$ and each component of $\mathbf{c}_k$, and G-Wishart priors for the precision matrix $\boldsymbol{\Omega}^{(k)} \sim W_G(3,\mathbb{I}_p)$ conditional on the graph $G^{(k)}$,  for $k=1,\ldots, K$. We let the MCMC chain run for 2 million iterations and discard the first 500k as burn-in.

\begin{figure}[!t]
	\centering
	\includegraphics[width=0.49\textwidth]{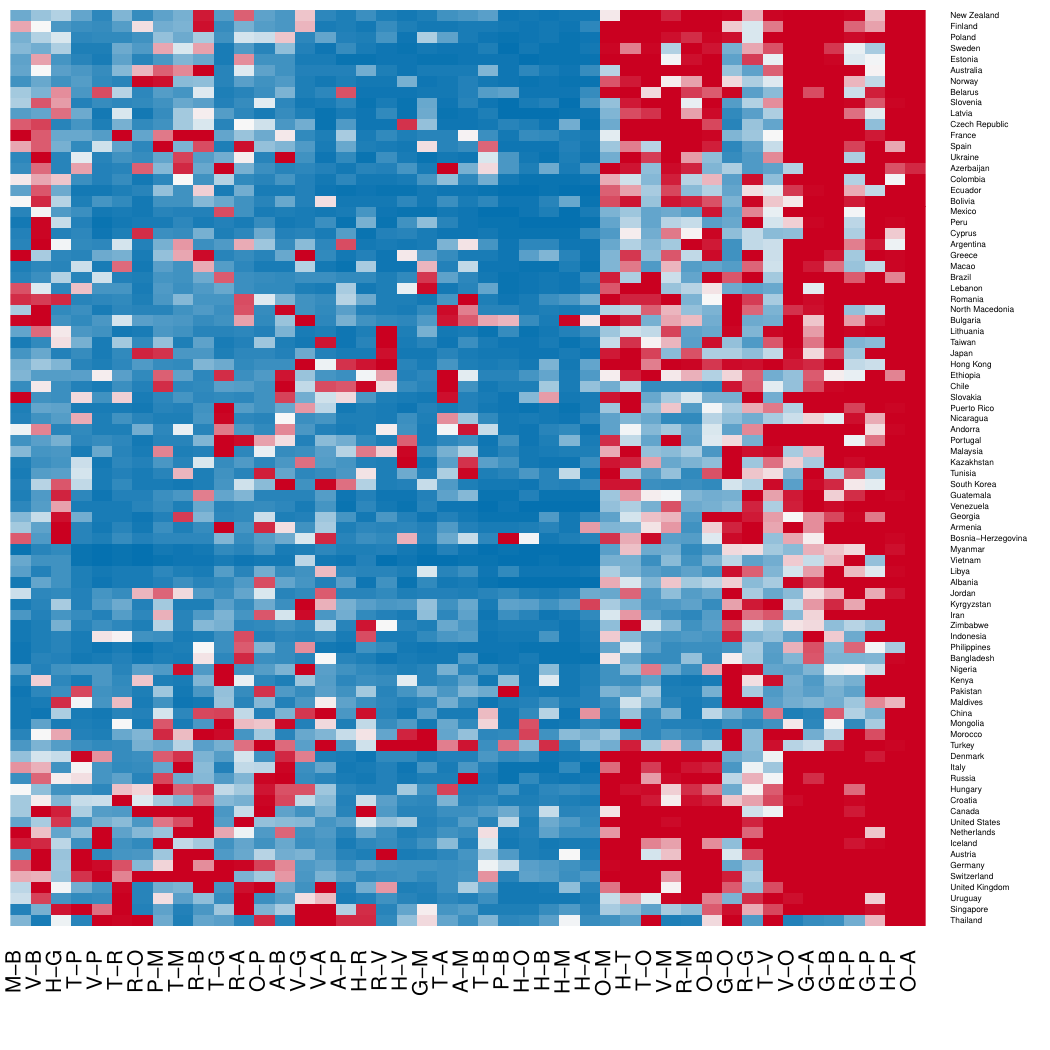}
\caption{Heatmap of posterior edge probabilities for each country, ranging from 0 (blue) to 1 (red). Results from the random graphical model with country-specific intercepts and latent space (Equation \ref{eq:modelLS}).}
\label{fig:modelLS1}
\end{figure}

\begin{figure*}[!t]
\centering
\begin{tabular}{cc}
 \includegraphics[width=0.47\textwidth]{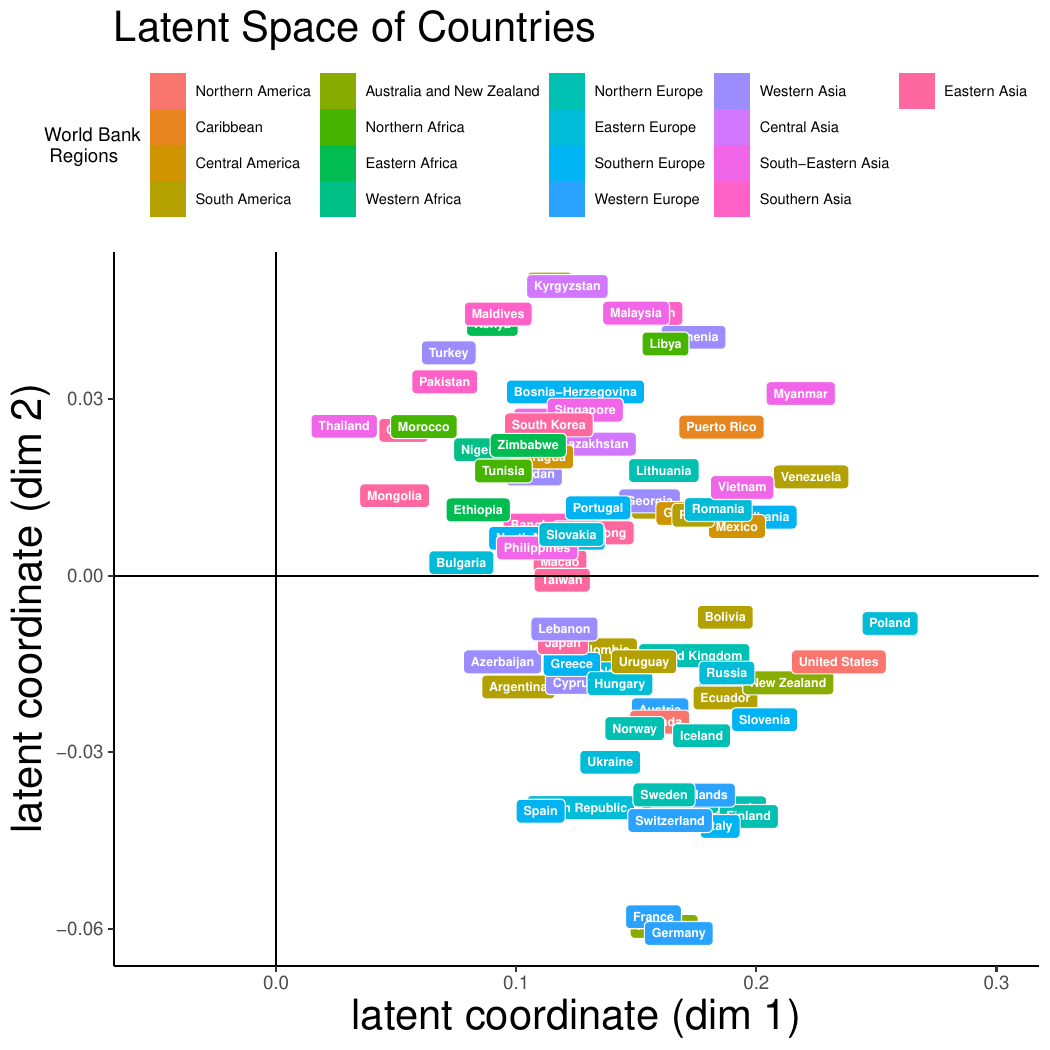} &
 \includegraphics[width=0.47\textwidth]{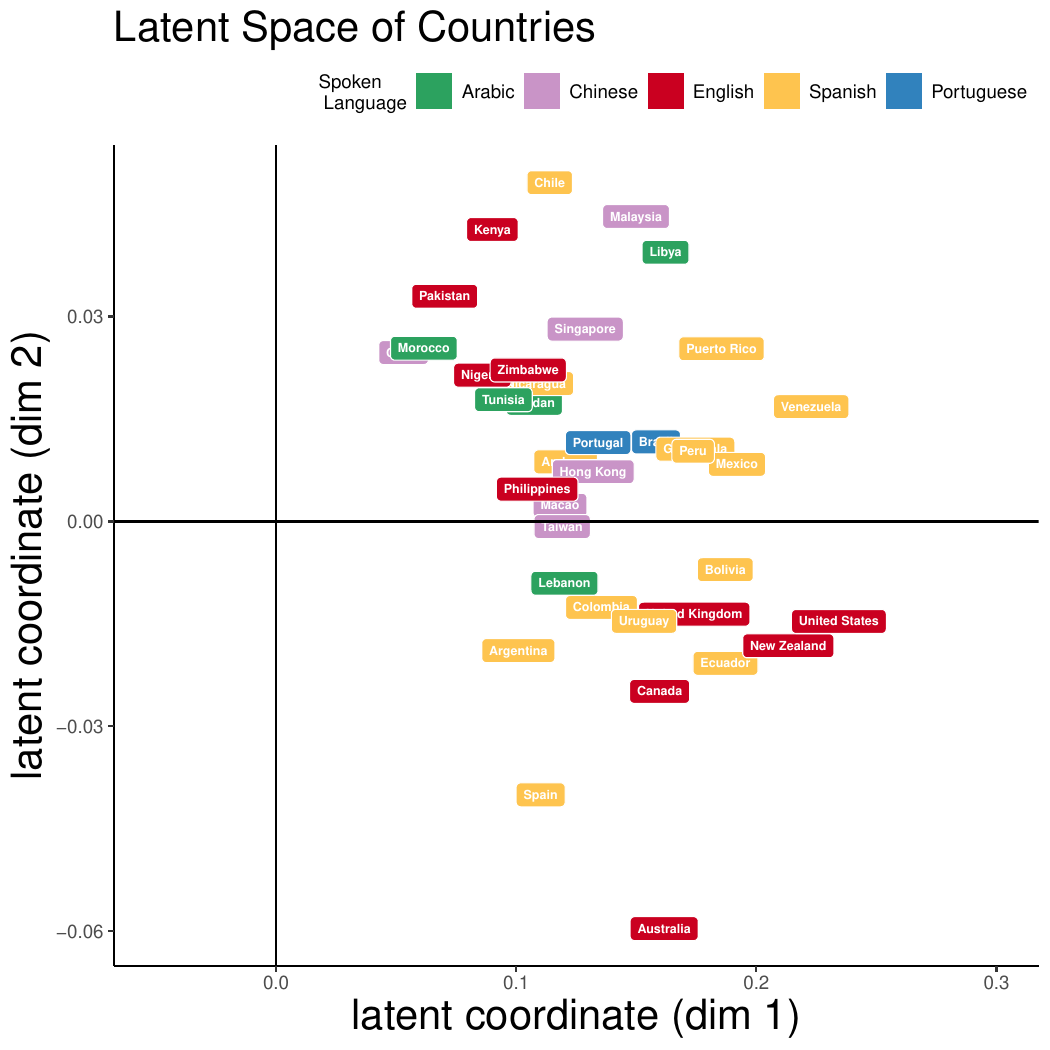}
\end{tabular}
\caption{Fitted latent space model with super-imposed colours associated to (left) World Bank regions and (right) official spoken languages for a selection of main languages, with Chinese including both Cantonese and Mandarin (right). Results from the random graphical model with country-specific intercepts and latent space (Equation \ref{eq:modelLS}).}
 \label{fig:modelLS}
\end{figure*}


Figure \ref{fig:modelLS1} is a heatmap of posterior probabilities for each edge (column) of the inferred cultural networks for each country (row). Since probabilities close to 0 are in blue and close to 1 in red, the plot shows how: 1) the networks tend to be in general quite sparse; 2) there is little uncertainty in the recovery of the network, with probabilities close to the two extremes; 3) there is a high degree of structural similarities between the networks, with connections like homosexuality and abortion (last column) present in all networks, while other topological structures are more local, i.e., specific to groups of countries. 

The fitted latent space is shown in Figure \ref{fig:modelLS} (left). Each country is placed in this plot at their estimated location, given by the posterior mean of the corresponding $\mathbf{c}_k$ parameter vector. The plot shows, first of all, a wide spread of countries, i.e., a large cultural heterogeneity. Secondly, countries that are located close to each other in this space, such as Sweden and Finland, have similar cultural networks, which we can see also from the posterior edge probabilities in Figure \ref{fig:modelLS1}, while countries that are located far away on this space, like Germany and Kyrgyzstan, are associated to structurally different networks. 

\begin{table}
	\caption{DIC values of four competing random graphical models, with a different formulation of the random graph generative model.\label{tb:DIC}}
	\centering
\resizebox{\textwidth}{!}{
	\fbox{%
		\begin{tabular}{*{3}{lcc}}
			random graph model & parameters & DIC \\ \hline
			country-specific intercepts \citep{debenedictis23} & $\alpha_k$  & 3,116,506\\
			country-specific intercepts + latent space (equation (\ref{eq:modelLS})) & $\alpha_k$, $\mathbf{c}_k$ & 3,115,201\\
			country-specific intercepts + proximity measures & $\alpha_k$, $\boldsymbol\beta$ & 3,095,245\\
			country-specific intercepts + proximity measures + latent space (equation (\ref{eq:latentprobit}))& $\alpha_k$, $\boldsymbol\beta$, $\mathbf{c}_k$ & 3,100,461\\
	\end{tabular}}}
\end{table}

The informativeness of the latent space can be assessed statistically by comparing this model to the model by \cite{debenedictis23}. In this framework, individual networks for each country with no structural sharing, as in \cite{debenedictis23}, can be obtained using a random graphical model with a generative graph defined only by country-specific intercepts $\alpha_k$. The Deviance Information Criterion (DIC) is used to compare the two models \citep{gelman14}. Denoting with $D(\Theta) = -2 \log L(\Theta)$ the deviance of a model with parameters $\Theta$ and likelihood $L(\Theta)$, the criterion is defined by
\[DIC = D(\hat\Theta) + 2\mbox{Var}(D(\Theta)),\]
where  $D(\hat\Theta)$ is the deviance evaluated at the mean posterior
estimate of the parameters, while $\mbox{Var}(D(\Theta))$ is the variance of the deviance, which we evaluate on a random sample of 50 MCMC draws of $\Theta$. As shown in Table~\ref{tb:DIC}, the DIC of the model with country-specific intercepts and latent space is lower than the model with only country-specific intercepts, leading to the selection of the more complex latent space model.

Figure \ref{fig:modelLS} shows how the location of the countries in the latent space has an association to geographical and linguistic connections between the countries. In particular, the left plot shows how countries that are close to each other geographically tend to be located close to each other on the latent space. Indeed, one can recognise the clusters of European countries (in shades of blue), Asian countries (purple/pink), African countries (green) and Latin American countries (brown). However, geographical distance explains only part of the heterogeneity, with for example United States and New Zealand located close to each other in the latent space. The right plot shows a further partitioning in terms of spoken language, for a selection of countries associated to the five spoken languages. It appears clearly from this plot how linguistic proximity plays also a role in explaining cultural similarities. In view of this exploratory analysis, in the next section, we expand the model to include geographical and linguistic distances as potential drivers of cultural heterogeneity and thus to quantify their potential association to culture.

\subsection{Identifying drivers of cultural heterogeneity}
\label{drivers}
We now include in the model the four potential drivers of cultural heterogeneity described in Section \ref{sec:distances}, namely geographical proximity of two countries, the sharing of an official or a primary language, and being in the same continent. Thus, we now consider the full formulation of the model in (\ref{eq:latentprobit}), with $\boldsymbol\beta$ a 4-dimensional vector of parameters with components associated to the four proximity measures, respectively. We set relatively flat $N(0,10)$ priors also on these parameters.

\begin{figure*}[!tb]
\centering
 \includegraphics[width=.9\textwidth]{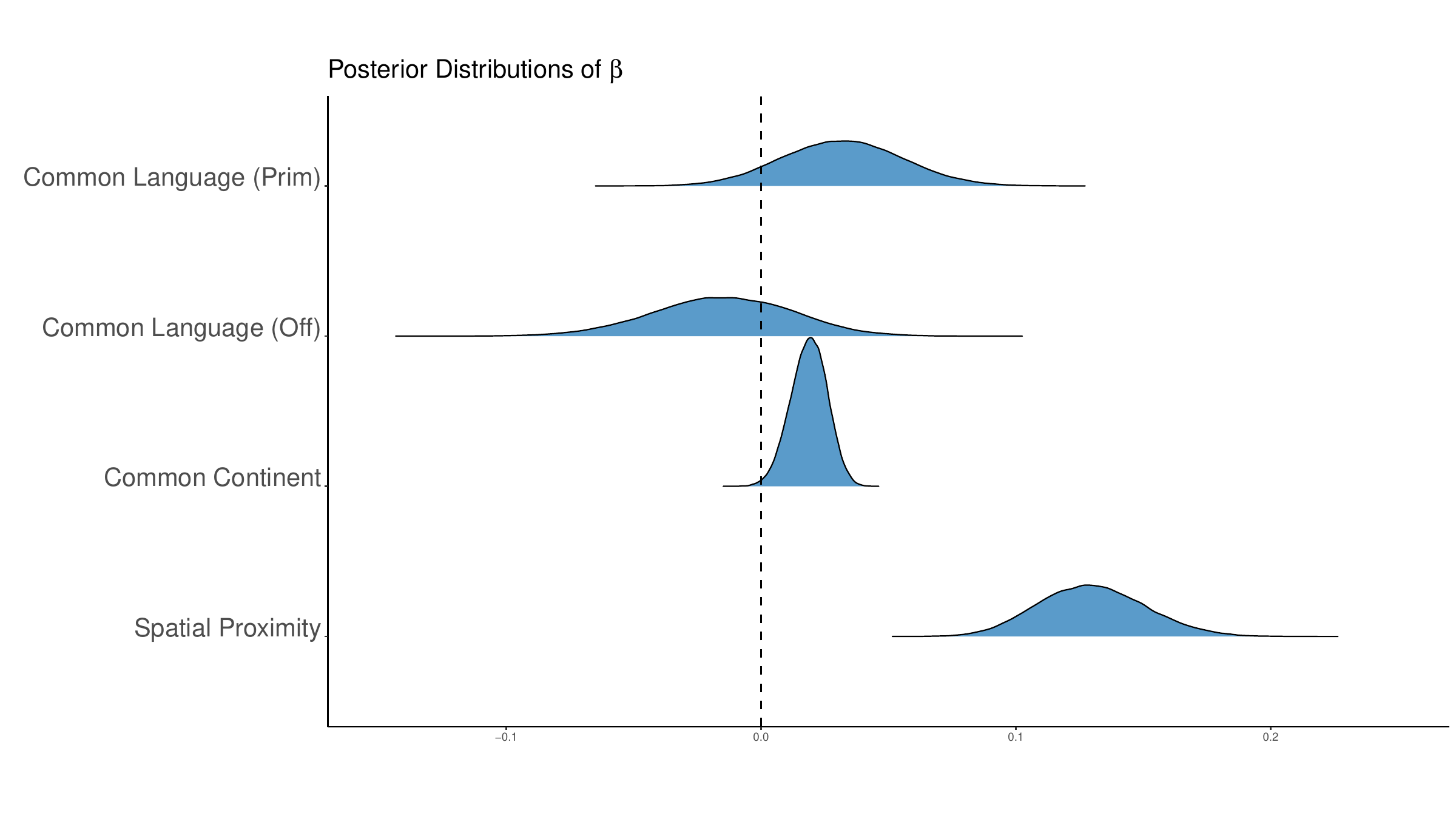}
\caption{Posterior distributions of $\boldsymbol\beta$ in model (\ref{eq:latentprobit}), associated to a selection of geographical and historical proximity measures.}
 \label{fig:distances}
\end{figure*}

Figure  \ref{fig:distances} displays the posterior distributions of $\boldsymbol\beta$. The results confirm a strong association between shared cultural values and the geography of the two countries, both in terms of geographical proximity and in terms of belonging the same continent. Sharing the same primary language also plays a role, often possibly via its connection to a history of colonisation between the two countries, while sharing the same official language is not significant, probably due to a correlation with the other language effect. 

\begin{figure*}[!tb]
\centering
\begin{tabular}{cc}
 \includegraphics[width=0.47\textwidth]{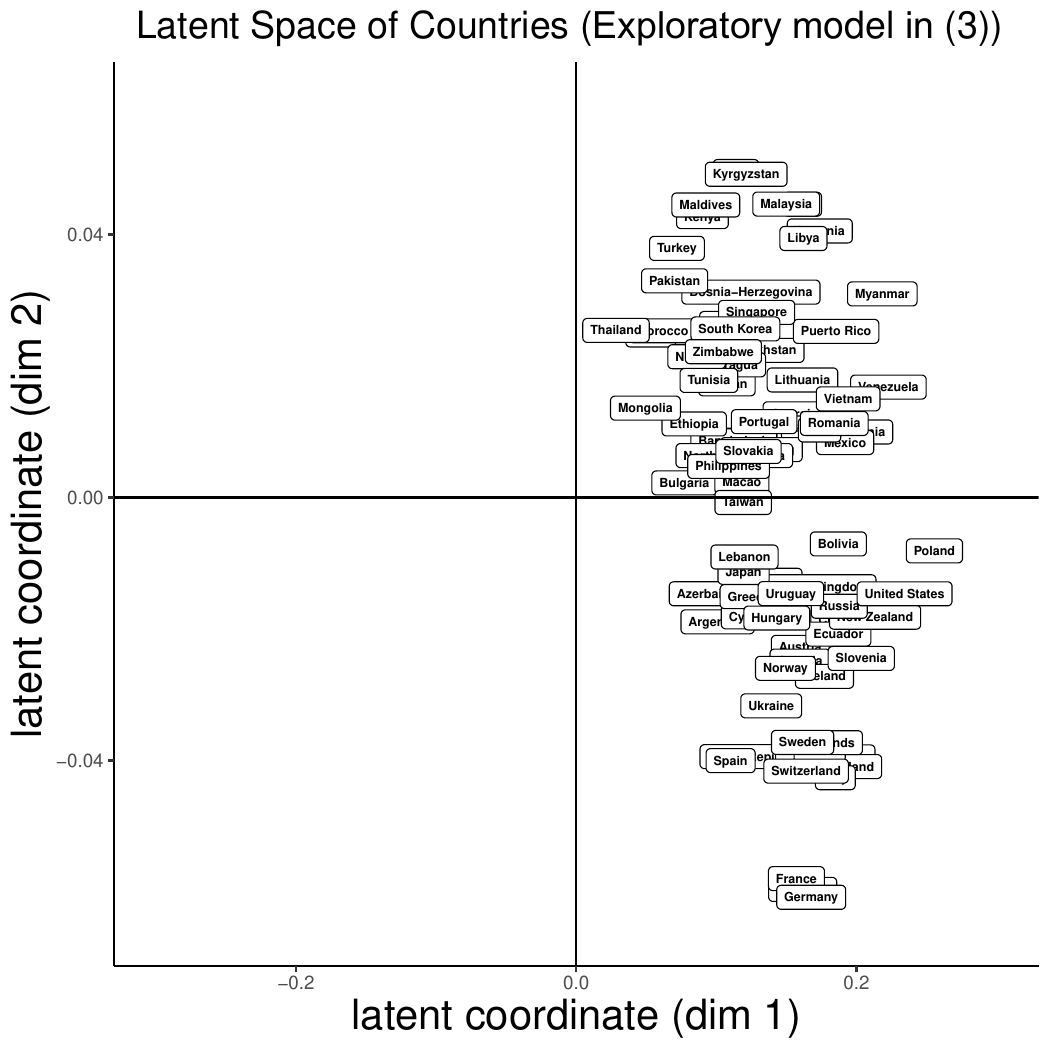} &
\includegraphics[width=0.47\textwidth]{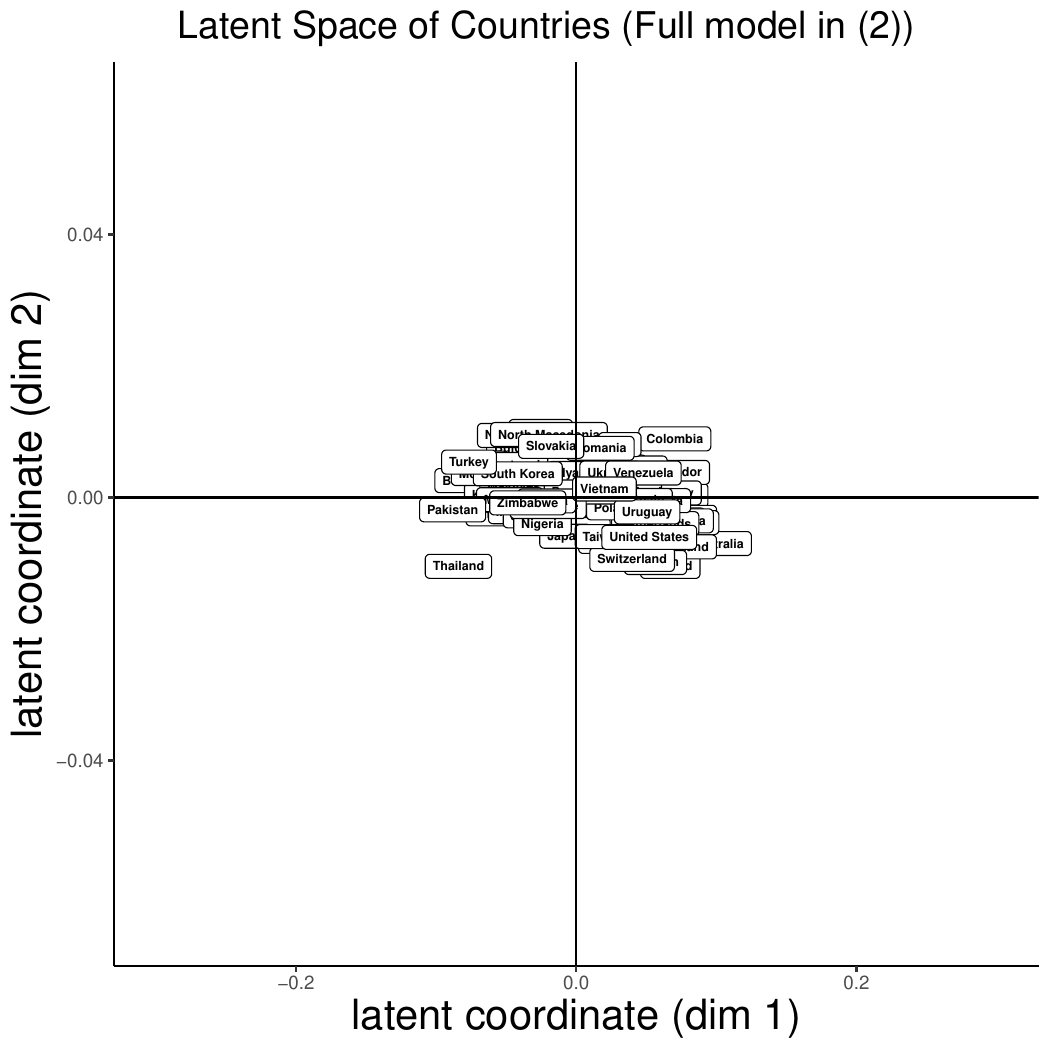}
\end{tabular}
\caption{Comparison on the same scale of the latent space in the exploratory model described in (\ref{eq:modelLS}) versus the residual latent space of the explanatory model described in (\ref{eq:latentprobit}) with country-specific intercepts and proximity measures. As can be seen, the suggested drivers capture most of the residual variation.}
 \label{fig:modelDLS}
\end{figure*}

As a further validation of the effect of the proximity measures on culture,  Figure \ref{fig:modelDLS} shows how the residual latent space of this model (right) is now not as informative as the latent space from the earlier exploratory model (left). This is supported by the DIC comparison in Table \ref{tb:DIC}, which shows how the model with the lowest DIC is the one that includes country-specific intercepts and proximity measures in the random graph generative process but excludes the latent space. This in fact turns out to be the best model out of all four models considered in this analysis.

\section{Conclusions}
\label{conclusion}
This paper has studied the extent of cross-country cultural heterogeneity by proposing an advanced graphical modelling approach that is able to integrate data at different levels and from different sources. The primary source of data is given by the responses to a selection of survey questions within each country. In particular, this paper considers a selection of 10 questions in relation to various cultural dimensions that have been answered in 84 countries around the world. The responses are both related to the marginal characteristics of the respondents, such as age and gender, as well as the dependence structure between the various cultural dimensions within each country. These two components are coupled together by country-specific Gaussian copula graphical models. Furthermore, these graphical models are modelled jointly via the inclusion of a random graph generative model that captures the socio-cultural similarity of countries. 

In our marginal models, we concentrated particularly on the demographic characteristics age and gender of the respondents. They carried significant importance in explaining within country differences of opinion on various cultural dimensions. It is essential to adjust for their effect on cultural views, as these can have an impact both on the estimation of the country-specific dependence structures and on the cross-country comparisons between samples with a different age-gender configurations. 

In our random graph model that generates jointly the cultural networks around the world, we concentrated on geographical and linguistic similarity measures between countries. The inclusion of these variables made the presence of a latent space in the random graph model superfluous. Our analysis showed, in fact, a large heterogeneity in cultural networks across the world, that for a large part is explained by geographical and linguistic proximity between countries.   

Compared to existing cross-country cultural studies, the proposed methodology allows to  1) adjust for marginal effects on cultural values, 2) learn cultural networks for each country, 3) identify cultural similarities across countries and their potential drivers.  The joint modelling of these three components, on the one hand, provides a fair ground for cross-country comparative studies and, on the other hand, leads to increased accuracy in parameter estimation, as similarities between cultures induce a sharing of information between the country-specific graphical models compared to separate analyses for each country.

The proposed framework lends itself easily to various extensions, both in terms of integrating new sources of data, such as alternative surveys, and of introducing new generative models. As for the latter, an interesting future direction would be to exploit all seven waves of the value survey data and to develop a dynamic version of the approach, in order to unveil the evolution of national culture and the potential cross-country influence on cultural changes.

\bibliographystyle{chicago}
\bibliography{references}

\end{document}